\documentstyle[twocolumn,prl,eqsecnum,aps,epsf]{revtex} % twocolumn format

\def\l{\ifmmode{\Lambda}\else{$\Lambda$}\fi}
\def\f{\ifmmode{\phi}\else{$\phi$}\fi}
\def\w{\ifmmode{\omega}\else{$\omega$}\fi}
\def\s{\ifmmode{\Sigma}\else{$\Sigma$}\fi}
\def\sn{\ifmmode{\Sigma^0}\else{$\Sigma^0$}\fi}
\def\sr{\ifmmode{\Sigma^{*}(1385)}\else{$\Sigma^{*}(1385)$}\fi}
\def\pt{\ifmmode{p_{\rm T}}\else{$p_{\rm T}$}\fi}
\def\xf{\ifmmode{x_{\rm Feynman}}\else{$x_{\rm Feynman}$}\fi}
\def\xfs{\ifmmode{x_{\rm F}}\else{$x_{\rm F}$}\fi}
\def\pr{\ifmmode{P_r}\else{$P_r$}\fi}
\def\pn{\ifmmode{P_N}\else{$P_N$}\fi}
\def\dnn{\ifmmode{D_{NN}}\else{$D_{NN}$}\fi}
\def\dns{\ifmmode{D_{NS}}\else{$D_{NS}$}\fi}
\def\ay{\ifmmode{A_y}\else{$A_y$}\fi}
\def\an{\ifmmode{A_N}\else{$A_N$}\fi}
\def\pkl{$\vec{p}p\to pK^+\vec{\Lambda}$}

\def\gev{{\rm GeV}}
\def\gevc{\ifmmode{\,{\rm GeV}/c}\else{$\,{\rm GeV}/c$}\fi}
\def\gevcc{\ifmmode{\,{\rm GeV}/c^2}\else{$\,{\rm GeV}/c^2$}\fi}

\def\mevc{\ifmmode{\,{\rm MeV}/c}\else{$\,{\rm MeV}/c$}\fi}
\def\mevcc{\ifmmode{\,{\rm MeV}/c^2}\else{$\,{\rm MeV}/c^2$}\fi}
\def\pvec{\ifmmode{\vec{p}}\else{$\vec{p}$}\fi}
\def\nvec{\ifmmode{\vec{n}}\else{$\vec{n}$}\fi}
\def\lesim{{$\lower 2pt\hbox{$\scriptstyle <$}\atop\raise4pt\hbox{$\scriptstyle\sim$}$}}
\def\grsim{{$\lower2pt\hbox{$\scriptstyle >$}\atop\raise4pt\hbox{$\scriptstyle\sim$}$}}
\begin{document}
\draft
\wideabs{
\title{Spin Transfer in Exclusive \l{} Production from
$\pvec p$ Collisions at 3.67\gevc}
\author{F.~Balestra$^1$, Y.~Bedfer$^2$, R.~Bertini$^{1,2}$,
	L.~C.~Bland$^3$, A.~Brenschede$^{4,a}$, F.~Brochard$^2$,
	M.~P.~Bussa$^1$, V.~Chalyshev$^5$, Seonho~Choi$^{3,b}$, 
        M.~Debowski$^{6,c}$,
	M.~Dzemidzic$^{3,d}$, J.-Cl.~Faivre$^2$, I.~V.~Falomkin$^5$,
	L.~Fava$^1$, L.~Ferrero$^1$, J.~Foryciarz$^{6,7,e}$, V.~Frolov$^5$,
	R.~Garfagnini$^1$, D.~Gill$^8$, A.~Grasso$^1$, S.~Heinz$^{1,2}$,
	V.~V.~Ivanov$^5$, W.~W.~Jacobs$^3$, W.~K\"uhn$^4$, A.~Maggiora$^1$,
	M.~Maggiora$^1$, A.~Manara$^{1,3}$, D.~Panzieri$^1$, H.-W.~Pfaff$^4$,
	G.~Piragino$^1$, G.~B.~Pontecorvo$^{1,5}$, A.~Popov$^5$,
	J.~Ritman$^4$, P.~Salabura$^6$,
	F.~Tosello$^1$, S.~E.~Vigdor$^3$,
	G.~Zosi$^1$\\(DISTO Collaboration)}
\address{$1${\rm )} Dipartimento di Fisica ``A. Avogadro'' and INFN, Torino, Italy}
\address{$2${\rm )} Laboratoire National Saturne, CEA Saclay, France}
\address{$3${\rm )} Indiana University Cyclotron Facility, Bloomington, Indiana, U.S.A.}
\address{$4${\rm )} II. Physikalisches Institut, Univ. Gie\ss{}en, Germany}
\address{$5${\rm )} JINR, Dubna, Russia}
\address{$6${\rm )} M.~Smoluchowski Institute of Physics, Jagellonian
University, Krak\'{o}w, Poland}
\address{$7${\rm )} H.~Niewodniczanski Institute of Nuclear Physics,
Krak\'{o}w, Poland}
\address{$8${\rm )} TRIUMF, Vancouver, Canada}
%\address{$e$) Gesellschaft f\"ur Schwerionenforschung, Darmstadt, Germany}
%\address{$j$) Institut f\"ur Kernphysik - University of Frankfurt}
\date{\today}
\maketitle
\begin{abstract}
We report the first polarization transfer measurements for
exclusive hyperon production reactions.
The normal spin transfer coefficient \dnn{} for 
\pkl{} is large and negative for forward \l{} production
 at a beam momentum of 3.67\gevc,
a result qualitatively consistent with expectations for a 
mechanism dominated by kaon-exchange and rescattering. The sign of
\dnn{} is opposite to that observed in the fragmentation regime for
inclusive \l{} production at much higher energies.
\end{abstract}
}% end of \wideabs{
\narrowtext
One of the primary challenges in the study of hadronic interactions is
the identification of the most effective degrees of freedom for their
theoretical description: in what regimes is it essential to consider
explicitly the interactions of the underlying quarks and gluons, where
is meson-exchange an efficient alternative, and what
(possibly hybrid) descriptions can be used in the transition between
these two regimes? One may hope to elucidate such a transition by
examining the evolution  of a given class of reactions over a very
broad energy range.

A promising case study is offered by polarization measurements in the
production of hyperons in proton-proton
collisions. Here, a large body of existing data\cite{Hell96} for
inclusive reactions appears to span the transition regime: over a wide
c.m. energy range ($5<\sqrt{s}<60\,\gev$) and at transverse momentum 
transfers $\pt \approx 1 $ to $ 3 \,\gevc$, they reveal stable polarization
effects that are too large to be understood in a partonic framework
based on perturbative quantum chromodynamics. These results
have been interpreted in a variety of simple {\it ad hoc} models, based on
diagrams involving either constituent
quarks\cite{Hell96,Degr85,Boro97} or Reggeized meson
exchange\cite{Soff91}. However, data at higher and lower energies 
are needed to anchor the existing interpretations in
regions where the relevant degrees of freedom ought to be
more clear. 

We report here the first hyperon
polarization results at energies closer to threshold, needed to
determine the viability of meson-exchange approaches where they have
worked successfully in reproducing cross section
measurements\cite{Lage91}.
The present experiment is the first to combine polarized beam
with {\em exclusive} hyperon production kinematics.  The exclusivity
(achieved without polarized beam in \cite{Feli96}) removes
significant ambiguities as to whether observed $\Lambda$'s result from direct
production or from heavier hyperon decay\cite{Bert89}.  Our 
measurement has
sufficient resolution to separate the $\pvec p\to
pK^+ \Lambda$, $pK^+ \Sigma^0(1192)$, and $pK^+ \Sigma^*(1385)$
reactions.

Polarized beam allows measurements of
the reaction polarization [$\pn(Y)$ for hyperon $Y$] to be
supplemented by the analyzing power (\an{}) and the spin transfer
coefficient (\dnn{}).
The subscripts $N$ label quantization axes
normal to the production
plane formed by the beam proton and the hyperon momentum
vectors.  $\dnn$ measures the fraction of the normal beam
polarization component transferred to the hyperon.
We concentrate here on the \dnn{} results because they are sizable
and subject to simple interpretation within either meson-exchange or
constituent quark reaction models\cite{Boro97,Vigd96}.
% Comparison of these new polarization measurements with the
% higher-energy systematics can reveal whether a unified description,
% in terms of a common set of degrees of freedom, is possible all the
% way from threshold to $\sqrt{s}$ = 60 GeV.

\begin{figure}
\epsfxsize=88mm
\epsfysize=61mm
\epsfclipon
\epsffile[47 294 565 652]{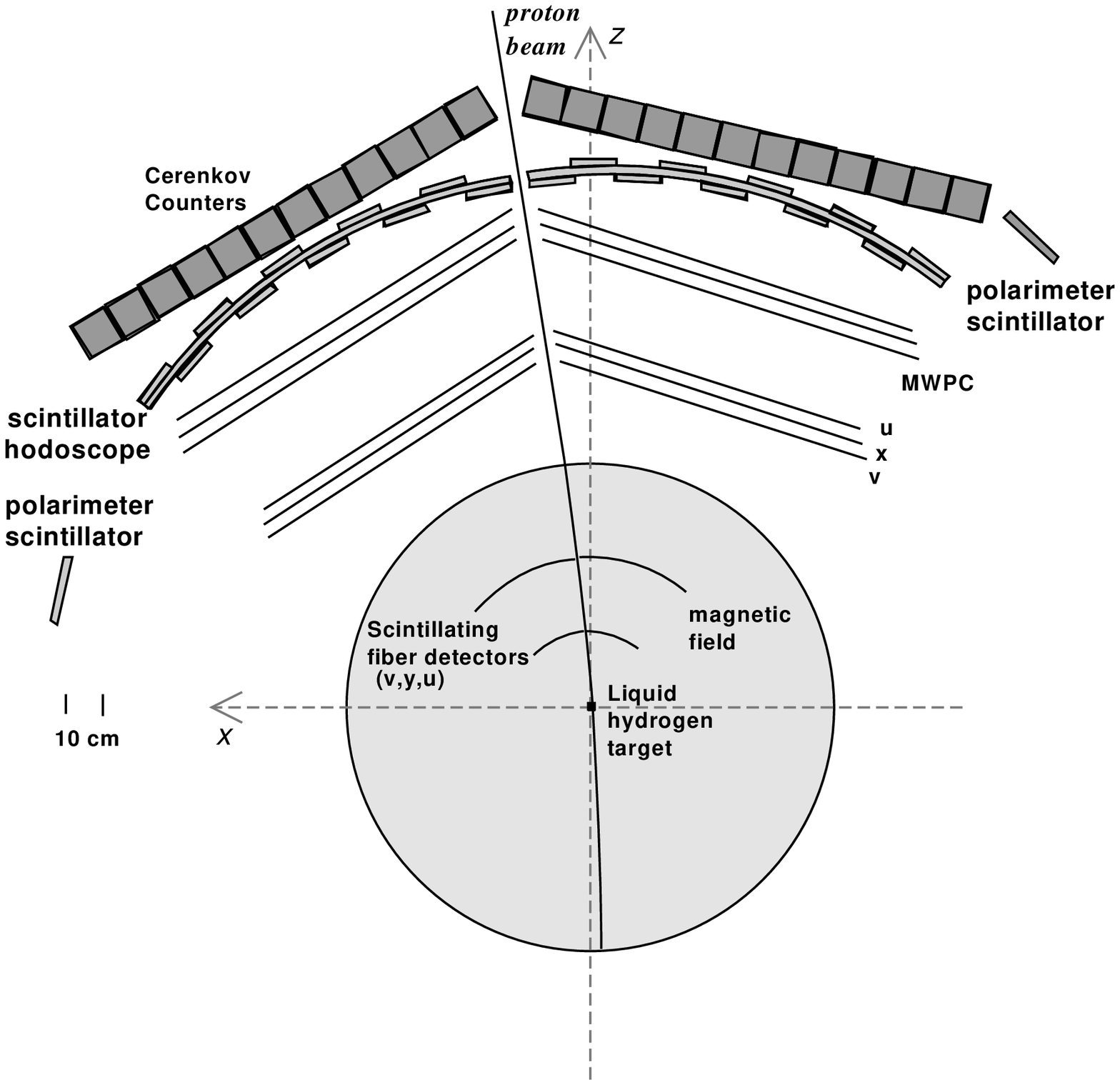}
\epsfclipoff
\caption{Schematic view of DISTO detector elements.}
\label{fig:disto_setup}
\end{figure}

Figure~\ref{fig:disto_setup} shows a schematic view of the 
experimental setup \cite{distonim} which was installed at the Laboratoire National
Saturne at Saclay.  A polarized proton beam of
3.67 $\gevc$ was scattered from a liquid
hydrogen target placed at the 
center of a large gap (40~cm) dipole magnet.  Multiple
charged reaction products were
detected in counters arranged in two arms mounted symmetrically on
both sides of the curving beam trajectory.  The particles were
tracked with scintillating fiber (SF) detectors and multi-wire
proportional chambers
(MWPC's), each composed of 3 stereo planes with elements aligned
vertically ($x$) or horizontally ($y$), and at $\pm 45^\circ$ ($u,v$)
to the vertical.  Particle
multiplicities, energy loss, and flight times were measured with a plastic
scintillator hodoscope comprising 6 horizontal and 10
vertical elements per arm.  Particle velocities (for $\beta > 0.75$)
were determined with an array of 12
water \v{C}erenkov counters mounted vertically at the rear of each
arm. In the laboratory frame, the detectors covered an
angular range from $\approx 2^\circ$ to 
$\approx 48^\circ$ horizontally, on both sides of the beam, and
$\pm15.5^\circ$ vertically. 
The large acceptance improves sensitivity in determining the
polarization of the \l-particles from the parity-violating
asymmetry in their weak decay to $p \pi^-$.

Two additional vertical
scintillator slabs were added beyond the angular range of the
hodoscope to detect recoil protons as part of a $\pvec p$
elastic scattering coincidence event stream used for on-line
monitoring of the beam polarization (P$_b$). 
The asymmetries, $\epsilon =$ P$_b$A$_N \sim 0.1$,
measured with this polarimeter were stable over time and
determined with a statistical precision {}$\sim \pm 0.001$  
during a typical two-week running period. The polarimeter A$_N$ was 
taken from the systematics of existing $\pvec p$ elastic 
scattering data \cite{Lehar}. 
The magnitude of the (vertical) beam polarization,
averaged over all production runs, is $0.73\pm 0.05$,
where the error reflects the normalization uncertainties in 
A$_N$ data in the 2-3 GeV energy range.
The beam spin direction was
flipped on alternate beam bursts to minimize 
time-dependent instrumental asymmetries.

The hyperon production trigger \cite{distonim} required at least three charged
tracks in
the final state, based on the multiplicity of fired hodoscope elements and
scintillating fibers.
%This trigger was used typically for 18 of each 20 successive beam
%bursts. For the remaining two bursts, we used a polarimeter trigger
%based on $pp$ elastic scattering angular correlations among hodoscope
%elements and polarimeter scintillators, selected by programmable
%lookup modules.
%The polarimeter yields were fed to scalers for every beam burst. 
%Alternatively,
%programmable lookup modules allowed appropriate selection of hodoscope
%and extra polarimeter slabs for the two-body elastic scattering used
%to monitor the polarization. Data were taken in ``ping-pong'' mode,
%with typically 18 bursts of hyperon production and 2 bursts of elastic
%trigger.
The beam intensity ($\sim 10^8$ protons/burst) and
trigger conditions were adjusted to limit the total data acquisition
dead time typically to $12$\%.  
%Details of the custom-designed data 
%acquisition and trigger hardware are presented in Ref.~\cite{ieee}.

Position information from the SF's, MWPC's, and
hodoscopes was used to reconstruct curved tracks through the
magnetic field and locate their intersection points. Pulse height
information from the hodoscope and \v{C}erenkov counters, together with
reconstructed momenta for each track, were used for particle
identification. Hyperon production candidates were selected
from events with four reconstructed tracks, consistent with
identification as $p$ and $K^+$ from a primary vertex within the
target volume and $p$ and $\pi^-$ from a decay vertex displaced by
$\geq 1$ cm.  The spectra presented here have been subjected only to
loose kinematic cuts, designed to yield minimal loss of good hyperon
production events.

\begin{figure}
\epsfxsize=85mm
\epsfysize=40mm
\centerline{\epsffile[28 308 580 571]{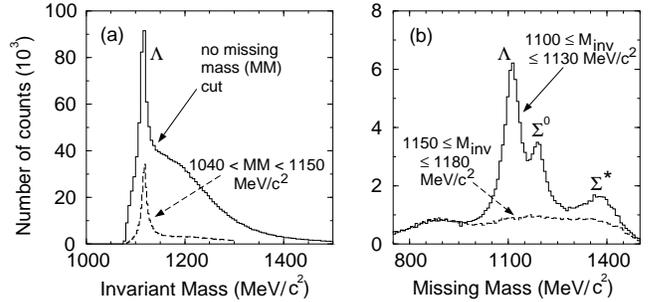}}
\caption{(a) Spectrum of $p\pi^-$ invariant mass at the decay vertex
(measured with resolution $\sigma \approx 5$ MeV/c$^2$) with (dashed)
and without (solid) a cut on the \l{} missing mass.  (b) Missing mass
spectrum ($\sigma \approx 25$ MeV/c$^2$) from primary $pK^+$ pairs
for $p \pi^-$ invariant mass in the range
1100--1130\mevcc{} (solid line), and 1150--1180\mevcc{} (dashed line).}
\label{fig:mass}
\end{figure}

Exclusive $\Lambda$ production events were identified as those with
both the invariant mass of the $p \pi^-$ pair ($M_{p\pi^-}$) and the 
missing mass ($\Delta M_{pK}$)
reconstructed from the $p K^+$ pair equal, within resolution, to the
$\Lambda$ mass.  The \l{} peak is seen clearly in both mass spectra in
Fig.~\ref{fig:mass}.  Two types of background under the \l{} missing
mass peak
must be removed to isolate the exclusive production:  (1) combinatoric
background ($p$ and $\pi^-$ not from \l{} decay) \cite{K0note}; 
(2) feedthrough from the adjacent
peak due to \sn{} production and decay to $\Lambda \gamma$. 
All other \l{} production channels, such as $pp \rightarrow
pK^+ \Sigma^*(1385),~pK^+ \Lambda \pi^0$, or $pK^0 \Lambda \pi^+$
(the latter contributing only via $\pi^+ - K^+$ misidentification), are too
far removed in missing mass to contribute appreciably under the direct
\l{} peak.

To infer the shape and magnitude of the combinatoric background in
the $\Delta M_{pK}$ spectrum we have selected, within each analyzed
kinematic bin, the event sample with $1150 \leq M_{p\pi^-} \leq     
1180$ \mevcc.  (As shown by the dashed spectrum in
Fig.~\ref{fig:mass}(a), the $M_{p\pi^-}$ background above the \l{} 
peak is quite flat when examined within a gate on the \l{} 
{\em missing} mass peak.) 
This sample is dominated by the abundant $pp\pi^+\pi^-$
events, which occasionally satisfy our loose hyperon production cuts
for $\pi^+$ momenta \grsim 1 GeV/c (where $\pi - K$ particle
discrimination is limited).  As indicated by the dashed curve in
Fig.~\ref{fig:mass}(b), this sample reproduces very well the non-physical
``hyperon'' events below 950 \mevcc, and accounts
for a fraction $f_{comb} = 0.12 $ to $ 0.31$ (depending on kinematic bin) 
of the area under the \l{} peak.

After subtracting the combinatoric background in each kinematic bin,
we have fitted the \l-\sn{} region with a sum of two Lorentzian
peaks. This peak shape reproduces well, the simulated spectra for 
pure \l{} or pure \sn{} production, when these are analyzed with
the same event reconstruction software as the real data.
The fits determine the  
relative enrichments [$f_{\Lambda,\Sigma}^i$
where $f_\Lambda^i + f_\Sigma^i = 1$] in \l{} vs. \sn{}
production of two event samples, within $\Delta M_{pK}$  
gates around the \l{} ($i=1$, 1040--1150 \mevcc) and \sn{} 
($i=2$, 1165--1240 \mevcc) peaks.  For example,
$f_\Sigma^1$ varies from 0.01 to 0.17 over the bins analyzed.

The polarization of the $\Lambda$'s is manifested by a fore-aft
(pseudoscalar)
asymmetry of the angular distribution of the daughter proton in the \l{} rest
frame, measured with respect to a spin quantization axis parallel to the
momentum vector product ${\bf
k}_{\rm beam}\times{\bf k}_\l$.  
By analyzing separately the event samples acquired with beam spin up
vs. down, and with ${\bf k}_\l$ to the left vs. right of ${\bf k}_{\rm
beam}$, one can separate the \l{} polarization into a component
independent of the beam spin (arising from the reaction
polarization \pn) and a component that reverses sign when the beam
spin is flipped (corresponding to \dnn) \cite{endnote}. 
The subtracted combinatoric
background is always statistically consistent with symmetry in the
``decay'' angular distribution, hence with \pn =\dnn =0, as
expected for the parity-{\em conserving} background processes. 

\vspace{2mm}

\begin{figure}
\epsfxsize=85mm
\epsfysize=70mm
\epsffile[10 45 555 540]{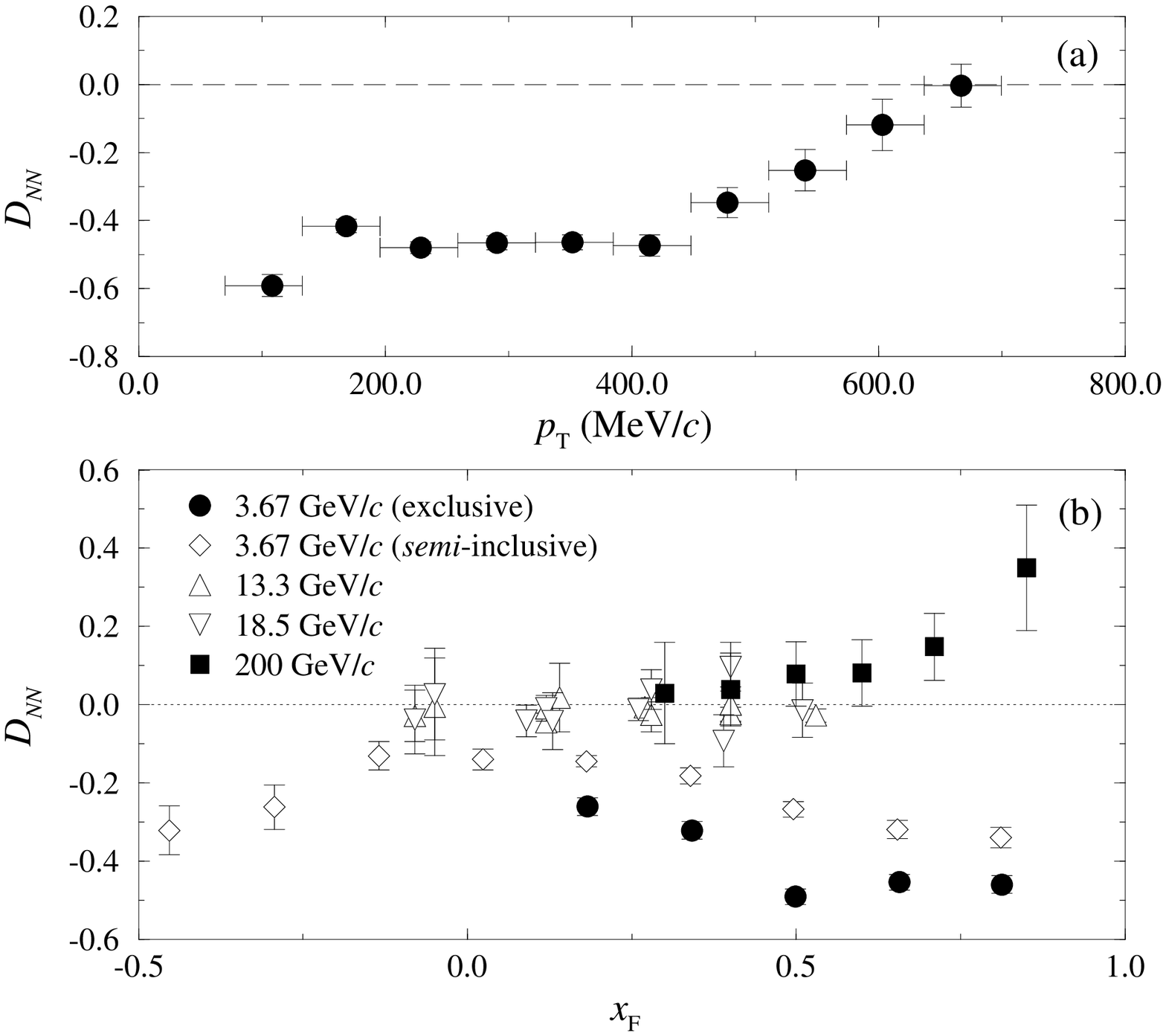}
\caption{(a) Measured \dnn{} values vs. transverse
momentum transfer for the {\em exclusive} \pkl{}
reaction. The horizontal error bars reflect the width of the \pt{}
bins analyzed. The vertical error bars reflect statistical
uncertainties only.
(b) \dnn{} as a function of \xfs{} for the present {\em exclusive}
\l{} production and for {\em inclusive} \l{}
production at various higher incident momenta from
Refs.~\protect\cite{Bonn87,Brav97}. Also shown for 
comparison are the {\em semi}-inclusive results from the
present data.}
\label{fig:dnn}
\end{figure}

For each $\Delta M_{pK}$ gate ${D_{NN}}^i$ is a linear combination,
$${D_{NN}}^i = f_\Lambda^i D_{NN}(\Lambda) +
f_\Sigma^i D_{NN}(\Sigma^0 \to \Lambda), ~i=1,2, $$
from which we extract the pure direct \l{} and $\Sigma^0 \to \Lambda$
results separately, using the fitted enrichments
$f_{\Lambda,\Sigma}^i$.  Here, 
$D_{NN}(\Sigma^0 \to \Lambda)$ represents the product of the
polarization transfers in the ${\vec p}p \rightarrow p K^+ 
{\vec \Sigma}^0$ production and the ${\vec \Sigma}^0 \rightarrow
{\vec \Lambda} \gamma$ decay.  The latter factor is typically
$- 1/3$.  The extracted $D_{NN}(\Lambda)$ values are always within
$\pm 0.04$ of the measured ${D_{NN}}^1$, and are stable with respect
to small shifts in the bounds of gates 1 and 2.

Figure~\ref{fig:dnn} shows the \dnn{} results for the exclusive \l{}
production as a function of both (a)
the transverse momentum transfer \pt{} (from $\pvec $ to $\vec{\l}$)
and (b) the \l's longitudinal momentum, expressed as a
fraction (\xfs) of its maximum kinematically allowed value. Exclusive
results are extracted only for $\xfs>0.1$ (the range included in
Fig.~\ref{fig:mass}), since at smaller \xfs{} the rising $p$ and $K^+$
momenta cause the missing mass resolution to deteriorate.
\dnn{} is large and negative ($\simeq-0.4$) over most of the
kinematic region. The negative sign signifies that the
component of the \l{} polarization that is correlated with the beam
spin is oriented {\em opposite} to the beam spin.

The error bars in Fig.~\ref{fig:dnn} include statistical, but not
systematic, uncertainties from the background subtraction and the
\l-\sn{} peak fitting.  We estimate the associated systematic
errors to be \lesim $\pm 0.02$ and $\pm 0.04$, respectively, allowing,
very conservatively, for shape changes that scale 
$f_{comb}$ by $\pm 20\%$ 
and $f_\Sigma^1$ by $\pm 75\%$.  The latter error is consistent with
the observed $D_{NN}(\Lambda) - {D_{NN}}^1$ differences. Instrumental 
asymmetries in detection efficiency do
not change when the beam spin is reversed, hence contribute errors
$< \pm 0.01$.  Overall scale uncertainties 
in \dnn{} arise from the present uncertainties in the beam polarimeter
calibration ($\pm 7\%$) and in the \l{} decay asymmetry ($\pm 2\%$
\cite{Part98}).  The systematic errors do not affect the striking qualitative
behavior observed for \dnn{} or the conclusions drawn below.

\vskip -0.3true cm
\begin{figure}
\epsfxsize=80mm
\epsfysize=50mm
\epsffile[30 468 542 761]{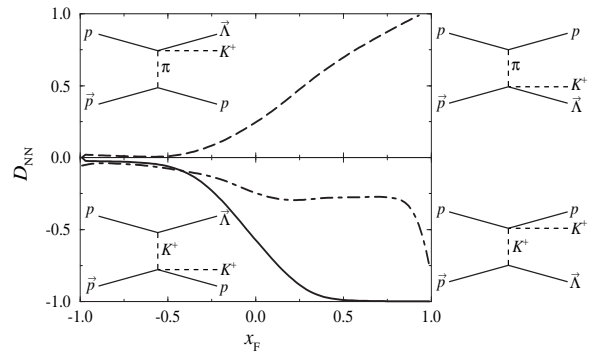}
\caption{Theoretical calculation \protect\cite{Lage91} of \dnn{} for
various {\em exclusive} \l{} production mechanisms: kaon-exchange (solid curve),
pion-exchange (dashed), or both combined with a \l-$p$ final state
interaction (dot-dashed). The Feynman diagrams indicate the dominant
exchange contributions for positive vs. negative \xfs{}. The
calculations are integrated over {\em all} phase space in other
kinematic variables.}
\label{fig:dnn_exp}
\end{figure}

The \dnn{} parameter is especially sensitive to the
hyperon production mechanism \cite{Vigd96}.
For example, in a meson-exchange framework, \dnn{} distinguishes
clearly between $\pi$- and $K$- exchange contributions. This is
illustrated in Fig.~\ref{fig:dnn_exp} by theoretical calculations employing the model
of Ref.~\cite{Lage91}. At large positive \xfs{}, \dnn{} is
maximally different for the 
two dominant contributions indicated in the figure. The difference 
arises from the different spin coupling at the vertices joining the
polarized proton to the hyperon.
To conserve angular momentum and parity, kaon emission at this 
vertex (bottom right diagram, Fig.~\ref{fig:dnn_exp}) causes a spin-flip,
%characterized by the operator {\boldmath$\sigma$}$\cdot {\bf q}$,
%where {\boldmath$\sigma$} represents the baryon spin and ${\bf q}$ the
%kaon momentum vector. Since ${\bf q}$ lies within the production
%plane, {\boldmath$\sigma$}$\cdot {\bf q}$ is purely off-diagonal in
%the spin projection normal to this plane, 
yielding $\dnn=-1$.

In contrast, the second mechanism (depicted in the top-right diagram of
Fig.~\ref{fig:dnn_exp}), where a virtual pion undergoes 
a strangeness-changing rescattering at the polarized proton vertex,
has a spin structure (${1\over 2}^+ + 0^-\to{1\over 2}^+ + 0^-$) for
which angular momentum and parity 
conservation require $D_{NN} = +1$.  (The latter holds rigorously for
the on-shell $\pi\pvec\to K\vec{\l}$ process,
but only approximately in this case, where the rescattering can have
an out-of-plane component of momentum transfer.)
The predicted \dnn{} tends toward zero for both mechanisms at the more
negative
\xfs{}, since the hyperon is then connected preferentially to the {\em
un}polarized (target) proton. Quark-line diagrams
lead to still different expectations for $D_{NN}$ \cite{Vigd96}.  

The large negative values observed for \dnn{} at $\xfs>0$ in
Fig.~\ref{fig:dnn}(b) and comparison with Fig.~\ref{fig:dnn_exp}
can thus be interpreted in a meson-exchange
framework to suggest kaon-exchange dominance. The full theoretical
calculation (dot-dashed curve in Fig.~\ref{fig:dnn_exp}, not yet
folded with the experimental acceptance) is
qualitatively similar to the measurements. 

How do the present results compare to polarization transfers measured
previously\cite{Bonn87,Brav97} for {\em inclusive} \l{} production at
higher energies? To facilitate this comparison, we include in
Fig.~\ref{fig:dnn}(b) \dnn{}
results for the {\em semi}-inclusive sample obtained by summing the
present data (after subtracting combinatoric background) over all 
missing masses above 1000\mevcc. Inclusion of
the \sn{} and \sr{} yields makes \dnn{} only slightly
less negative than our exclusive \l{} results at $\xfs > 0$.  
The present data clearly differ in sign from inclusive \dnn{} results
obtained at 200\gevc{} for large \xfs.

Although these different data sets
% in Fig.~\ref{fig:dnn}(b)
encompass quite different ranges of \pt{}, the dominant sign
of \dnn{} does not change with \pt{} at either the present or the highest
energies. The signs of the polarization observables, including \dnn{}
and \pn{}, for
inclusive \l{} production at 200\gevc{} \cite{Brav97} have been
successfully interpreted in a simple constituent quark model 
in Ref.~\cite{Boro97}.
The present results are not naturally explained within this quark model.

In summary, we have reported the first polarization transfer
measurements for an exclusive hyperon production reaction, \pkl. 
The large negative values observed for \dnn{} at positive \xfs{} 
represent a robust qualitative behavior that suggests a
%regimes where the outgoing \l{} carries most of the incident polarized
%proton's longitudinal momentum, are what one would expect for a
production mechanism dominated by kaon-exchange, and that differs markedly
from high-energy observations for inclusive hyperon production. 
Detailed comparison of meson-exchange \cite{Lage91} and other model 
calculations with the data will be important for a more refined
understanding.  
The addition of the present results to the existing
database for hyperon production affords an opportunity to track
%The present \dnn{} results are quite different from those 
%for inclusive hyperon production at higher energies, 
%affording an opportunity to track 
the evolution with energy in the relevant degrees of freedom.

This work has been supported in part by the following agencies:
CEA-DSM, CNRS-IN2P3, GSI, INFN, KBN (2 P03B 117 10 and 2 P03B 115 15)
and NSF. We thank J.-M. Laget for making his calculation
available and for useful discussions. We acknowledge the
excellent support provided 
by the SATURNE II accelerator and technical staff in
completing this program.


\begin{references}
\bibitem{K0note} $K^0$ decay contributions to the invariant mass
spectrum are eliminated by the complete $p$ {\it vs.} $\pi^+$ particle
discrimination.  This is confirmed by the absence of a $K_S^0$ peak when
we reanalyze events, treating the decay $p$ as a $\pi^+$.
\end{references}
\end{document}